\documentclass[aps,pra,floatfix,oneocolumn]{revtex4}

\usepackage{amsmath}
\usepackage{graphicx}
\usepackage{epstopdf}

\newcommand{\bra}[1]{\mbox{$\langle #1|$}}
\newcommand{\ket}[1]{\mbox{$|#1\rangle$}}

\begin{document}

\title{Effects of frequency correlation in linear optical entangling gates operated with independent photons}
\author{M. Barbieri}
\affiliation{Centre for Quantum Computer Technology, Department of Physics, University of Queensland, QLD 4072, Brisbane, Australia.}

\begin{abstract}
Bose-Einstein coalescence of independent photons at the surface of a beam splitter is the physical process that allows linear optical quantum gates to be built. When distinct parametric down-conversion events are used as an independent photon source, distinguishability arises form the energy correlation of each photon with its twin.  We derive upper bound for the entanglement which can be generated under these conditions.
\end{abstract}
\maketitle

\section{Introduction}

Interference of photons emitted by independent sources is one of the most intriguing features of the quantum theory of light and represents a remarkable departure from classical electromagnetism \cite{mandel,paul}. This phenomenon is easily understood if the radiation is treated in terms of creation and annihilation operators, rather than by invoking the superposition principle for light waves. This purely quantum effect represents the key feature for entanglement generation needed to build linear optical quantum gates \cite{klm}. These schemes rely on Hong-Ou-Mandel (HOM) effect, which is the observed coalescence of photons on the outputs of a beam splitter  when they arrive simultaneously on the two input ports \cite{HOM}. Since this effect is due to the Bose-Einstein statistics of photons, a classical wave description is inadequate to give an exhaustive picture of the phenomenon. In the long term, these gates promise to become the basic constituents of quantum computers, either in networked architecture \cite{klm, nc}, or adopting one-way quantum computation approach in order to build cluster states \cite{briegel,mike,owexp,clusterpan,clusterpino}. 

Correct functioning of these devices requires true single photon states. Due to the difficulty of generating such non classical states \cite{source1, source2}, to date parametric down conversion (PDC) has been widely used \cite{qptcnot, cnot1,  cnot2,  cnot3}, where high frequency pump photons are non-deterministically converted to pairs of frequency entangled daughter photons in a passage through a non-linear crystal. Independent photons can be produced via two independent PDC processes \cite{panint,int,till}: of the two photons generated in each down conversion event, one is detected to herald the presence of its twin which is subsequently sent to the non-classical interferometer. 

It is important to distinguish between imperfections intrinsic to the gate architecture and those which come from non-ideality of the soure \cite{till}.  PDC processes  is represented generate two photon states in the form $\ket{\psi}{{=}}\left(I+\epsilon a^\dag_1 a^\dag_2+\epsilon^2 (a^\dag_1)^2 (a^\dag_2)^2+o(\epsilon^3)\right)\ket{0}$, with $|\epsilon|<<1$. Higher order terms can not be discriminated without photon number resolving detectors, which are not widely available. Crucially, the frequency correlations of the interfering photons with their twins can introduce distinguishability and, consequently, {\em welcher weg} (which path) information. The presence of such knowledge intrinsically reduces the visibility of  the interference and the level of entanglement of the state generated by the quantum gate \cite{goedel, kwiat, fabio}. Many efforts have been devoted in designing PDC sources with engineered spectral correlations \cite{uren, carrasco}.

In the present paper, we give a theoretical analysis of the maximal entanglement which can be produced operating a linear optical gate with independent PDC photons. This is an important diagnostic for quantum gate engineering, in which we wish to know the contribution of the imperfections of the source to the error budget \cite{till}.  Note that a first detailed sketch of such a calculation can be found in Ref. \cite{eccala, eccala1}, where the effect of frequency correlation on entanglement swapping and multi-photon entanglement generation is studied. The results presented here represent a more general and detailed approach.

\section{Nonclassical interference of independent PDC photons. }

Our picture begins with the theoretical treatment of down conversion in Ref.\cite{rubin}, restricting, for the sake of clarity, to the paraxial approximation and considering the pumping with a plane wave beam in mode locking regime. Specifically, we want to calculate the optimal visibility in experiments of interference of two {\em single photons}, each one coming from a frequency entangled pair, given the spectra of the pump fileds. The interferometric apparatus is the one shown in Fig. 1:  photons 1/ 2, and  3/4 are pairwise produced in two distinct down-conversion processes. Consequently, photons 1 and 4 are directly coupled to the detection to provide the trigger signal, while photons 2 and 3 interfere on a beam splitter with transmittance $T$ and reflectance $R$. 

\begin{figure}[!htb]
\begin{center}
\includegraphics[ scale=.4, bb=100 370 500 700,clip]{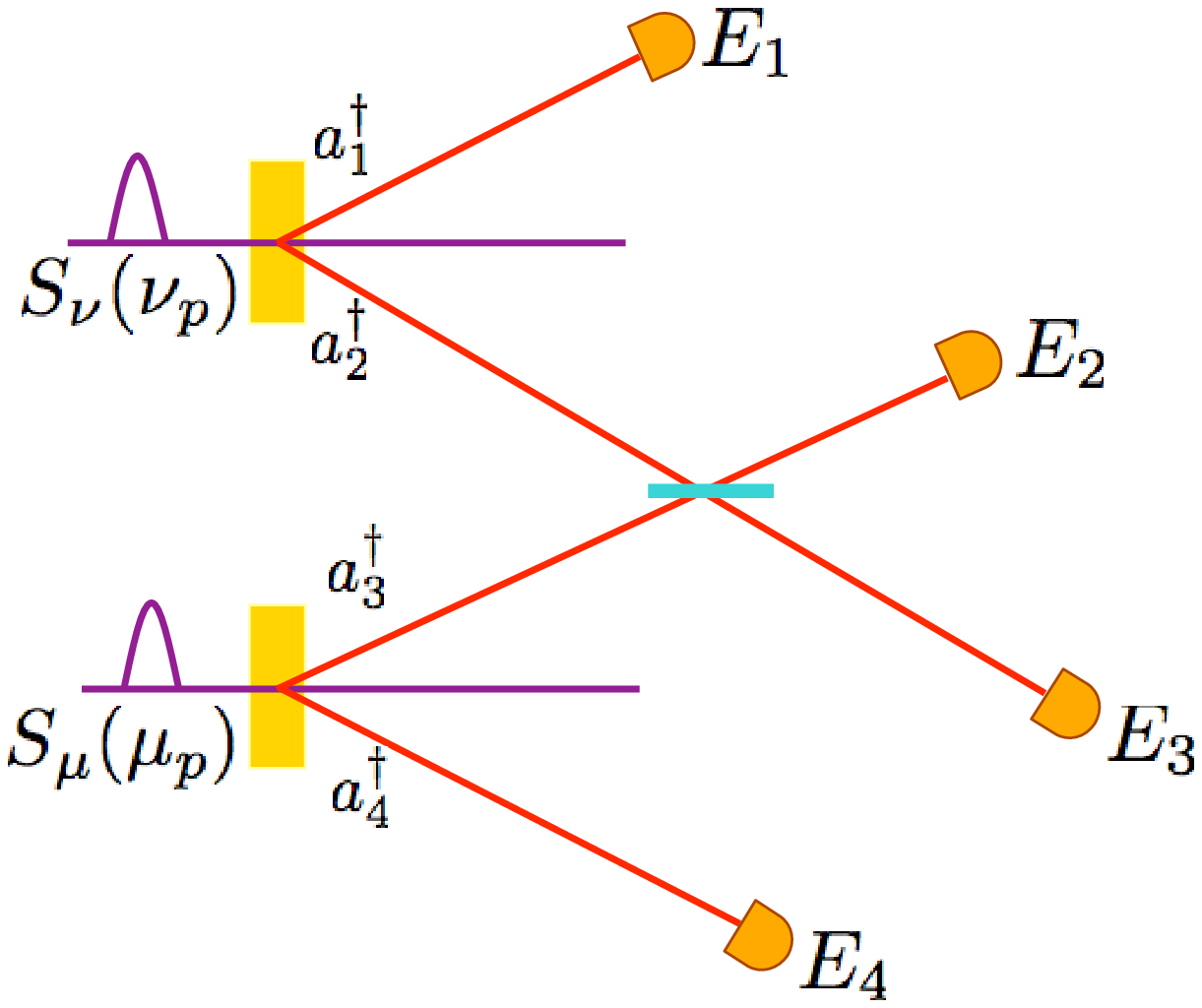}
\caption{(Color online) Schematic Hong-Ou-Mandel interferometer for independent photons. $S_\nu(\nu_p)$ and $S_\mu(\mu_p)$ represent the pump spectra for the upper and lower beam respectively. These produce of a photon pair on the modes $a_1^\dag$ and $a_2^\dag$, $a_3^\dag$ and $a_4^\dag$, respectively. Measurement of a photon in the $j{-}$th detector is described by the destruction operator $E_j$. }
\end{center}
\end{figure}

Consider, for instance, the PDC event occurring in the upper crystal in Fig.1 . The pump beam, traveling on the $z$ direction, has a frequency spectrum $S_\nu(\nu_p){{=}}s_\nu(\nu_p)e^{i\phi_\nu(\nu_p)}$ around a central value $\Omega_p$; $s_\nu(\nu)$ describes the distribution of the frequencies and $\phi_\nu(\nu_p)$ their phase relation with respect the central wavelength. In a single event, the pump generates of two fields at frequencies $\omega_1$ and $\omega_2$, with wavevectors $k_1$ and $k_2$, respectively. In these conditions, the two-photon wavefunction is given by,
\begin{equation}
\begin{aligned}
\label{wf}
\ket{\psi_{12}}{=}\int d\,\omega_{p}\,d\omega_1\,d\omega_2\, \, \delta(\omega_p-\omega_1-\omega_2)S_\nu(\omega_p-\Omega_p)a^\dag_1(\omega_1)a^\dag_2(\omega_2)\ket{0},
\end{aligned}
\end{equation}
up to a normalization factor. Effects due to the finite size of the crystal are neglected. In the other crystal, a similar process occur; the second pump beam is chosen to have the same central frequency $\Omega_p$, and a spectrum $S_\mu(\mu_p){{=}}s_\mu(\mu_p)e^{i\phi_\mu(\mu_p)}$.

The operator describing the photon annihilation due to its measurement is,
\begin{equation}
E_j{=}\int d\,\omega' e^{-4\ln2\left(\frac{\omega'-\Omega}{\sigma}\right)^2} a_j'(\omega '),
\end{equation}
where $j{{=}}1,2,3,4$ and $a'_j(\omega')$ represents the annihilation operator on mode $j$ at the frequency $\omega'$. In the expression above we adopt gaussian frequency filters, with central wavelength $\Omega$ and full width at half maximum (FWHM) $\sigma$ in front of the detectors: these define the coherence time of the photons. We choose the same (degenerate) frequency $\Omega{=}\Omega_p/2$ and the same FWHM for the four filters, since experiments are usually run near these conditions. Specifically, the measured fields at the time $t$ are given by the annihilation operators:
\begin{align}
&a'_1(\omega'){=}e^{-i\omega' t^0_1}a_1(\omega')\\
&a'_2(\omega'){=}e^{-i\omega' t^0_2}(Ta_2(\omega')+Re^{i\omega' \tau}a_3(\omega'))\\
&a'_3(\omega'){=}e^{-i\omega' t^0_3}(T^*a_3(\omega')-R^*e^{-i\omega' \tau}a_2(\omega'))\\
&a'_4(\omega'){=}e^{-i\omega' t^0_4}a_4(\omega'),
\end{align}
where $t^0_i{=}t_i-\ell_i$/c, $\ell_i$ optical path of $i-$th photon, c is the speed of light, and $\tau$ is the optical delay between the two arms \cite{loudon}.
The fourfold coincidence rate is,
\begin{equation}
C_4(\tau){\propto}\int dt \left|\bra{0} \
\prod_{j=1}^4 E_j\ket{\psi_{12}\otimes\psi_{34}}\right|^2,
\end{equation} 
where $dt$ is shorthand for the differential $dt_1dt_2dt_3dt_4$ \cite{glauber}. This integral is taken over the coincidence window we choose in our experiment; since typically this is much larger than the coherence time of the photons, and of the duration of the pump pulse, we can take integrals on infinite intervals  \cite{note}.  The calculation is then reduced in evaluating the amplitude, 
\begin{equation}
A(t_1,t_2,t_3,t_4;\tau){{=}}\bra{0}{\prod_{j=1}^4 E_j}\ket{\psi_{12}\otimes\psi_{34}}. 
\end{equation}
We use the coordinate transform
$t^\nu_+{=}\frac{t^0_1+t^0_2}{2}, t^\mu_+{=}\frac{t^0_3+t^0_4}{2}, t^\nu_-{=}t^0_1-t^0_2, t^\mu_-{=}t^0_4-t^0_3$. We also introduce  frequency detuning for the PDC: $\nu_i{=}\omega_i-\Omega$. Notice that $\nu_1+\nu_2{=}\nu_p$, $\nu_3+\nu_4{=}\mu_p$. Finally, it is found to be more convenient to express the amplitude in terms of the differences $\nu_-{=}\nu_1-\nu_2$, $\mu_-{=}\nu_4-\nu_3$, thus,
\begin{widetext}
\begin{equation}
\begin{aligned}
&A(t_+^\nu,t_-^\nu,t_+^\mu,t_-^\mu;\tau){=}\int d\nu_p \,d\mu _p\,d\nu_-d\mu_- \, S_\nu(\nu_p)e^{-\left(\frac{2\ln2}{\sigma^2}\right)\nu_p^2} S_\mu(\mu_p)e^{-\left(\frac{2\ln2}{\sigma^2}\right)\mu_p^2} e^{-\left(\frac{2\ln2}{\sigma^2}\right)(\nu_-^2+\mu_-^2)}\\
&\left\{|T|^2\left(e^{-i\nu_pt^\nu_+} e^{-i\mu_pt^\mu_+} e^{-i\nu_-t^\nu_-/2} e^{-i\mu_-t^\mu_-/2} \right)-|R|^2\left(e^{-i\frac{\nu_p}{2}(t^\nu_++t^\mu_++\frac{t^\nu_--t^\mu_-}{2}-\tau)}e^{-i\frac{\mu_p}{2}(t^\nu_++t^\mu_+-\frac{t^\nu_--t^\mu_-}{2}+\tau)}\right.\right. \times \\ &\left. \left.e^{-i\frac{\nu_-}{2}(t^\nu_+-t^\mu_++\frac{t^\nu_-+t^\mu_-}{2}-\tau)} e^{-i\frac{\mu_-}{2}(t^\mu_+-t^\nu_++\frac{t^\nu_-+t^\mu_-}{2}+\tau)}\right)\right\}.
\label{Anew}
\end{aligned}
\end{equation}
\end{widetext}
We neglect a global phase and normalization constants which are unnecessary for our calculations, since we will deal exclusively with ratios. If the integration over the frequencies is carried out we arrive to the expression:
\begin{widetext}
\begin{equation}
\begin{aligned}
& A(t_+^\nu,t_-^\nu,t_+^\mu,t_-^\mu;\tau){=} |T|^2\left( F_\nu(t_+^\nu) F_\mu(t_+^\mu) e^{-\delta_0^2(t_-^\nu)^2}e^{-\delta_0^2(t_-^\mu)^2}\right)+\\
&-|R|^2\left( F_\nu((t^\nu_++t^\mu_++\frac{t^\nu_--t^\mu_-}{2}-\tau)/2) F_\mu((t^\nu_++t^\mu_+-\frac{t^\nu_--t^\mu_-}{2}+\tau)/2)e^{-\delta_0^2(t^\nu_+-t^\mu_++\frac{t^\nu_-+t^\mu_-}{2}-\tau)^2}e^{-\delta_0^2(t^\mu_+-t^\nu_++\frac{t^\nu_-+t^\mu_-}{2}+\tau)^2}\right),
\label{AF}
\end{aligned}
\end{equation}
\end{widetext}
where $F_\nu(t)$ is given by the convolution 
\begin{equation}
F_\nu(t){=}\int dt' \tilde S_\nu(t')e^{-4\delta_0^2(t-t')^2},
\label{conv}
\end{equation}
$\delta_0^2{=}\sigma^2/32\ln2$, and $\tilde S_\nu(t)$ is the Fourier transform of the pump spectrum. $F_\mu(t)$ is defined in a similar way for the second pump.

The degree of indistinguishability in HOM interference is measured by its visibility, defined as $v{=}1{-}C_{m}/C_{M}$, with $C_m$ the coincidence rate at optimal temporal superposition of wavepackets (occurring at a delay $\tau_0$) and $C_M$ the rate in absence of interference. We thus can express $v$ in terms of the amplitude $A(t_+^\nu,t_-^\nu,t_+^\mu,t_-^\mu;\tau)$,

\begin{equation}
v{=}\frac{2|TR|^2}{|T|^4+|R|^4}\frac{Re[I_{S}(\tau_0)]}{I_N},
\label{visib}
\end{equation} 
where we have defined,
\begin{widetext}
\begin{equation}
I_S(\tau){=}\frac{\sqrt{\pi}}{2\delta_0}\int \,dx_1dx_2 dx_3 \, F_\nu^*(\frac{x_1+x_2}{2}) F_\mu^*(\frac{x_1-x_2}{2})F_\nu(\frac{x_1+x_3}{2})F_\mu(\frac{x_1-x_3}{2})e^{-2\delta_0^2(x_2-\tau)^2}e^{-2\delta_0^2(x_3+\tau)^2},
\label{zia}
\end{equation}
\begin{equation}
I_N{=}\frac{\pi}{2\delta_0^2}\int dx_1  \, |F_\nu(x_1)|^2 \int dx_2\, |F_\mu(x_2)|^2. 
\label{zio}
\end{equation}
\end{widetext}
A physical interpretation of  the integral $I_S(\tau)$ is not straightforward; nevertheless, it is evident that not only the time symmetry of each profile is involved, but also non trivial correlations between the two pump pulses. The fact that visibility is linked to the pump temporal profile in such a non trivial way, stresses the fact that this interference process is much more demanding than the one with correlated photon. The same pulse shape that gives a perfect HOM effect with frequency entangled photons, could be only partially suitable to reach high visibilities with independent photons.  
\begin{figure}[!]
\begin{center}
\includegraphics[scale=.6, bb=100 150 500 700,clip]{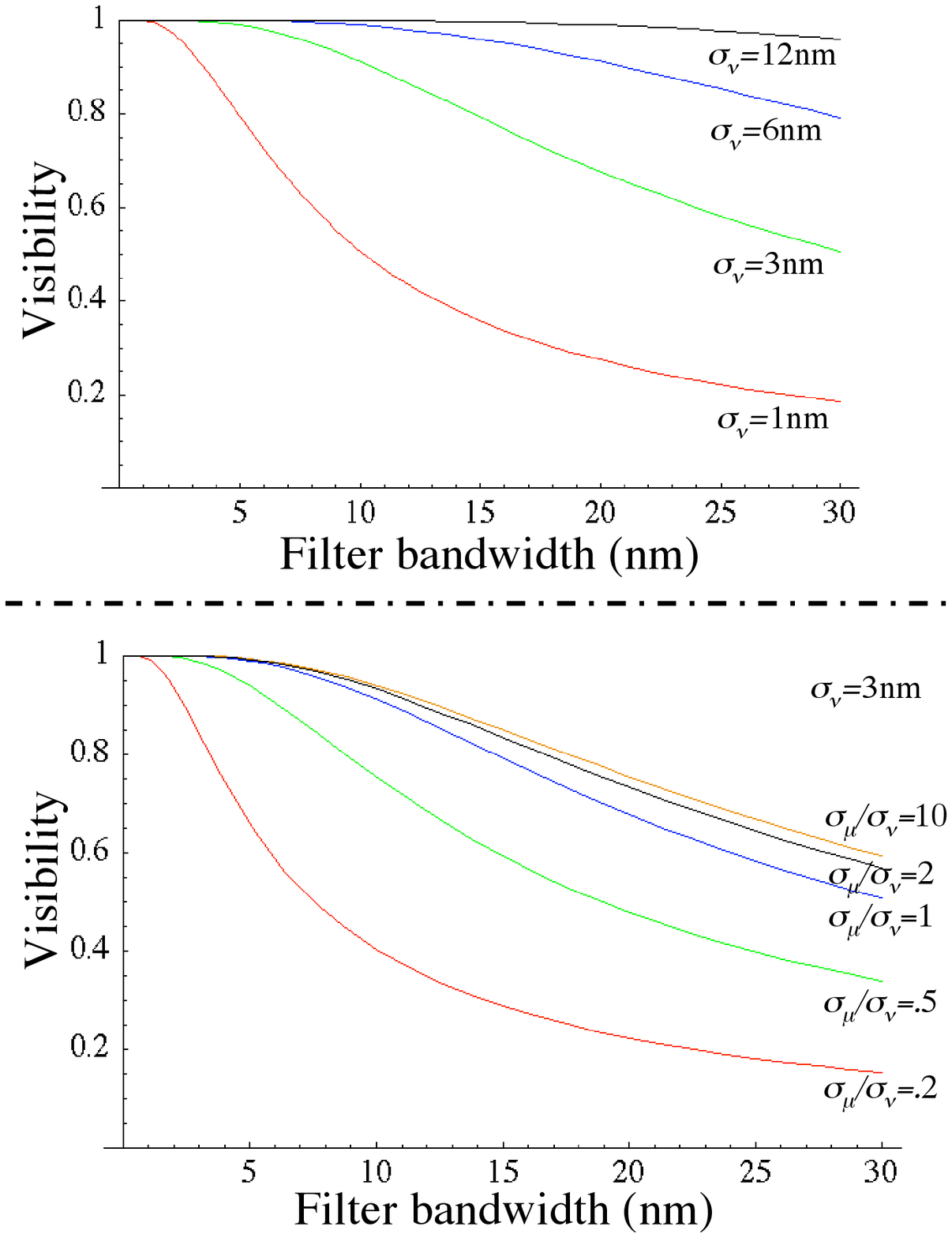}
\caption{(Color online) Visibility $v_0$ as a function of filter bandwidth $\sigma$ in absence of phase dispersion. Top: both pump spectra are gaussian profiles with the same FWHM $\sigma_\nu=\sigma_\mu$. Different plots correspond to different values of $\sigma_\nu$. Bottom:  the two gaussian spectra have different FWHM's. Plots relative to different ratios $\sigma_\mu/\sigma_\nu$ are shown for the case $\sigma_\nu{=}3$nm.}
\end{center}
\end{figure}
We can derive a simpler form for $I_S(0)$ in the case of infinitely large filters. In this case, we find
\begin{equation}
I_S(0)\propto\int dx_1 |\tilde S_\nu(x_1)|^2|\tilde S_\mu(x_1)|^2.
\label{chicco}
\end{equation}
By invoking Parseval's theorem, we can rewrite \eqref{chicco} in the frequency domain
\begin{equation}
I_S(0)\propto\int d\nu_0 |S_\nu*S_\mu(\nu_0)|^2,
\end{equation}
where the star denotes the convolution. More generally, a sufficient condition for achieving near perfect visibility is that  the convolutions have to be solutions of the integral equation:
\begin{equation}
\begin{aligned}
&F_\nu(x+y)F_\mu(x-y){=}\frac{4\delta_0}{\sqrt{\pi}}\int dz\,  e^{-8\delta^2_0(y^2+z^2)}F_\nu(x+z)F_\mu(x-z).
\label{15}
\end{aligned}
\end{equation}
It can be demonstrated that the condition of maximum visibility correspond to pump pulses which are much narrower in time than the coherence time imposed by the filters. In fact, the solutions of equation \eqref{15}  are $F_\nu(t){=}F_\mu(t){=}e^{-4\delta^2_0t^2}$; this are obtained by replacing $\tilde S(t)$ with a Dirac distribution. Substitution in \eqref{zia} and \eqref{zio} leads to the maximum visibility $v{=}1$ and Gaussian modulation of the interference pattern observed with correlated photons \cite{HOM}.  The considerations above indicate which physical process spoils the visibility \cite{eccala, eccala1}. When monochromatic pump beams are used, this imposes a very strict correlation to the frequency of the PDC photons. Hence, within the bandwidth of the filters some four photons events that do not interfere, since the wavelength of the two trigger photons may be used to reconstruct the path information. For instance, consider the event in which, after the beam splitter, photons 1 and 2 have exactly the frequency $\Omega_p/2$, and photons 3 and 4 are slightly detuned at the sides of the filter bandwidth. This can be realized by a unique configuration, viz. photons 1 and 2 from the first crystal and photons 3 and 4 from the second one. No event with the same wavelengths can be observed when photons 1 and 3 belong to the first pair and photons 2 and 4 to the second pair. When the frequency correlation is smoothed by adopting broader pump beams, such an information is erased and, consequently high visibility is restored. This descends from the fact that we introduced further four photon events which now cancel the ones previously non interfering. This process has been individuated as the responsible for degradation of polarization entanglement in similar experiments of multi-photon entanglement creation from  entangled pairs \cite{eccala, eccala1}. 

As an example, we consider a gaussian spectrum with chirping,
\begin{equation}
S_\nu(\nu_p){=}e^{-4\ln2\left(\frac{\nu_p}{\sigma_\nu}\right)^2}e^{i(\phi_1^{(\nu)}\nu_p+\phi_2^{(\nu)}\nu_p^2)},
\end{equation}
which gives in the convolution of eq. \eqref{conv},
\begin{equation}
F_\nu(t){=}\frac{1}{\sqrt{\gamma_\nu+4\delta_0^2}}e^{-\frac{4\delta_0^2\gamma_\nu(t-\phi_1^{(\nu)})^2}{\gamma_\nu+4\delta_0^2}},
\end{equation}	
where $\gamma_\nu{=}\frac{1}{4}\left(\frac{4\ln2}{\sigma_\nu^2}-i\phi^{(\nu)}_2\right)^{-1}$.  A similar expression holds for $F_\mu(t)$. Figure 2 show the maximum visibility $v_0{{=}}\frac{|T|^4+|R|^4}{2|T^*R|^2} v$ as a function of the filter bandwidth
without chirping. This is plotted for the condition of expected maximum visibility $\tau_0=0$. We find that high visibility is achieved with narrow filters and short duration hence large bandwidth pulses.  An imbalance in the bandwidth for the two pulses appears as a further source of distinguishability; Fig. 2 also show variations in the visibility curve as the ratio $\sigma_\mu/\sigma_\mu$ varies. it is clear that the control on this parameter doesn't need to be extremely accurate to achieve satisfactory values;  what primarily influences the visibility is the presence of a narrower bandwidth pump pulse, rather than the imbalance of the two pump bandwidths.

Phase dispersion effects of the second order are far less important. The second order coefficient $\phi_2^{(\nu)}$ is responsible for a broadening in time of $\tilde S_\nu(t)$ with respect the non chirped case. When narrow filters are used, its effects are negligible if the ratio $\phi_2^{(\nu)}/(4\ln2\sigma_\nu^{-2})$ remains of the order of few units. The linear coefficient $\phi_1^{(\nu)}$ describes a shift on the time axis, and can be use to described time jitter between the two pulses. When the pump beams are synchronous  ($\phi_1^{(\nu)}{=} \phi_1^{(\mu)}{=}0$), maximum visibility occurs at $\tau{=}0$; if the following pulses are separated by a jitter time $T_j$ (i.e. $\phi_1^{(\nu)}{=}0, \phi_1^{(\mu)}{=}T_j$) the interferometer is not working at its best delay, consequently, the visibility decreases to the value $v(T_j)$. An average over the total measurement time leads to the expected value $\int dT_j v(T_j)P(T_j) $, being $P(T_j)$ a proper distribution for the jitter \cite{int}. 

\section{Polarization entanglement generation via HOM interferometry.}
As said, HOM effect is the basic ingredient for building linear optical gates \cite{klm}. In particular, we can adopt interference on a partially polarizing beam splitter with $|T_H|^2=1/3$ ($|T_V|^2=1$) for the horizontal-$H$ (vertical-$V$) polarization to realize a non-deterministic controlled-sign gate \cite{cnot1, cnot2, cnot3}. The model depicted above can be utilized to calculate how much frequency entanglement affect the performance of such a gate when it is used for entanglement generation.
We describe entanglement generation via a partially polarizing  beamsplitter with 
$T_H=1/3$  and $T_V=1$, where extra losses on vertical modes are induced in order to obtain a correct balancement in the entangled state. In this architecture, vertical photons never meet at the surface of the beam splitter, while horizontal photons undergo HOM effect. This realizes the controlled interaction, together with post-selection via measurement. If a pair of photons with diagonal polarization $\ket{D}=\frac{1}{\sqrt{2}}\left(\ket{H}+\ket{V}\right)$ impinges on the beam splitter, by post-selecting coincidence events we generate the entangled state,
\begin{equation}
\ket{\varphi}=\frac{1}{\sqrt{2}}\left(\ket{VV}+\ket{VH}+\ket{HV}-\ket{HH}\right),
\end{equation}
since $\ket{HH}$ events are erased by the interference.
In the non ideal case, the polarization couples to the wavefunction of the pair in the time domain, leading to the following expression for the state of the pair,
\begin{equation}
\frac{1}{\sqrt{3+\epsilon^2}}\left(\ket{A}(\ket{VV}+\ket{HV}+\ket{VH})+\epsilon\ket{B}\ket{HH}\right),
\end{equation}
where $\ket{A}$ describes the time behavior without interference, and $\ket{B}$ is the time behavior with interference. Ideally, we would have $\ket{B}=-\ket{A}$, and $\epsilon=1$. The state can be rewritten as,
\begin{equation}
\frac{1}{\sqrt{3+\epsilon^2}}\left(\ket{A}(\ket{VV}+\ket{HV}+\ket{VH}+\epsilon c \ket{HH})+\epsilon \sqrt{1-|c|^2}\ket{A^\perp}\ket{HH}\right),
\end{equation}
where $c=\langle A \ket{B}$, and $\ket{A^\perp}$ is an orthogonal vector to $\ket{A}$. By tracing out the temporal part, the polarization density matrix is obtained,
\begin{equation}
\rho_\pi=\frac{1}{1+\epsilon^2}\left(\ket{\varphi'}\bra{\varphi'}+|\epsilon|^2(1-|c|^2)(\ket{HH}\bra{HH}\right),
\label{massimo}
\end{equation}
being $\ket{\varphi'}$ the non-normalized vector $\ket{VV}+\ket{HV}+\ket{VH}+\epsilon c \ket{HH}$.
In the case under investigation, we can identify up to the normalization,
\begin{align}
&\langle 0\ket{A}= F_\nu(t_+^\nu) F_\mu(t_+^\mu) e^{-\delta_0^2(t_-^\nu)^2}e^{-\delta_0^2(t_-^\mu)^2},\\
&\langle0\ket{B}=A(t_+^\nu,t_-^\nu,t_+^\mu,t_-^\mu;\tau).
\end{align}
Consequently, we find,
\begin{align}
&c=\frac{3}{\sqrt{5(1-v)}}\left(\frac{1}{3}-\frac{5}{6}v\right),\\
&\epsilon=\sqrt{5(1-v)},
\end{align} 
in the case in which $I_S(\tau)$ is a real function. It is straightforward to check that $c$ and $\epsilon$ reach their ideal values in correspondence of the optimal visibility $v_{id}{=}0.8$. A simple relation between the visibility and the degree of entanglement in the state \eqref{massimo}, quantified via the tangle \cite{wootters} is found,
\begin{equation}
T(v)= \frac{25v^2}{(8-5v)^2}.
\end{equation}
\begin{figure}[!htb]
\begin{center}
\includegraphics[ scale=.65, bb=100 423 500 700,clip]{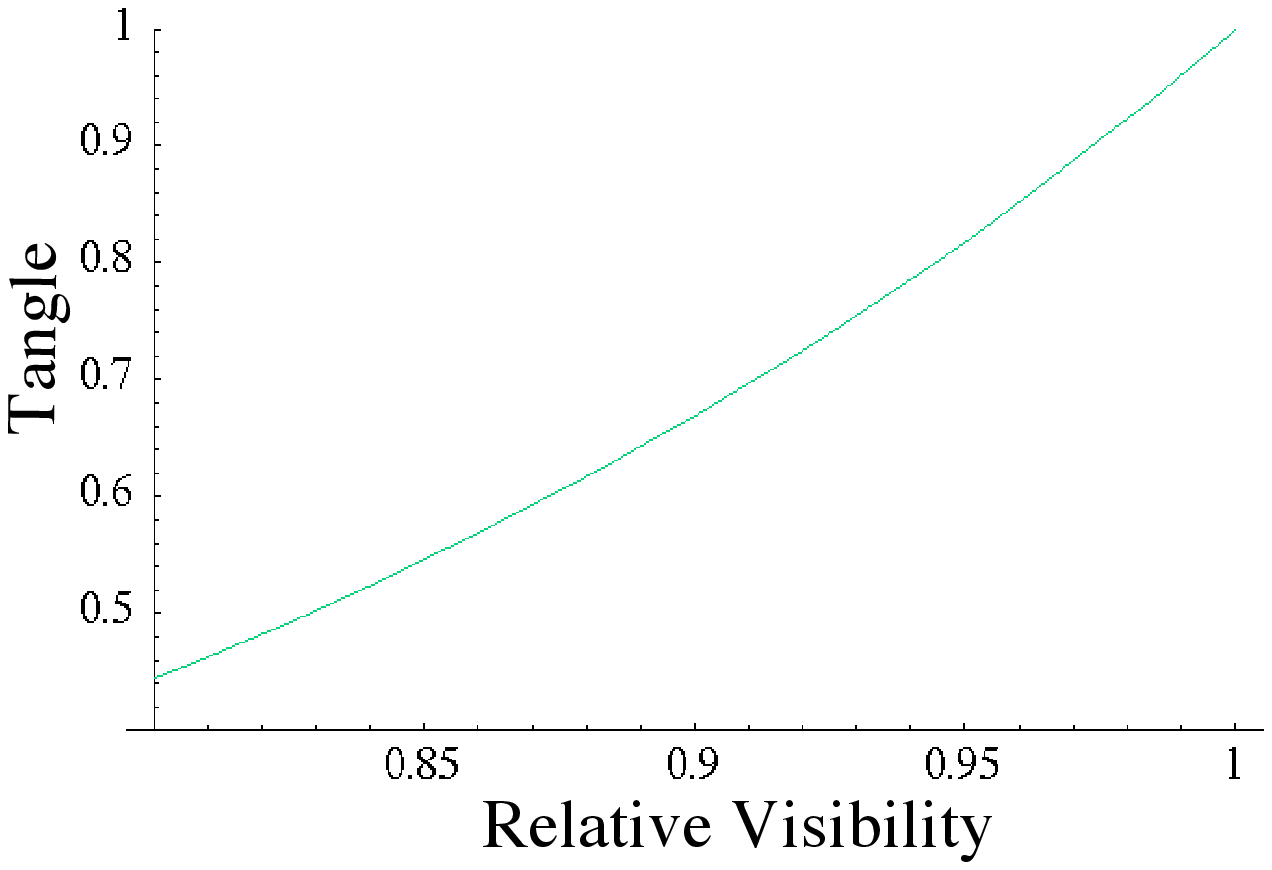}
\caption{(Color online) Plot of the tangle in the polarization state of eq. \eqref{massimo} as a function of the HOM visibility, normalized to its maximum value $v_{id}=0.8$}
\end{center}
\end{figure}

The plot in Fig. 3 shows the behavior of this curve as a function of the relative visibility $v/v_{id}$; it is clear that the degree of entanglement rapidly is degraded as the visibility decreases. Tangle remains well under the value 0.9, even for values of the visibility which are commonly considered more than satisfactory in  experiments. Thus, the maximum attainable entanglement strongly depends on the choice of the pump and the detection bandwidths, as shown in Fig. 4 for the case of Gaussian pumps.  A comparison of the figures indicates that tangle is a much stricter criterion in designing an experiment. 

\begin{figure}[!htb]
\includegraphics[width=0.49\columnwidth, bb=100 150 500 700,clip]{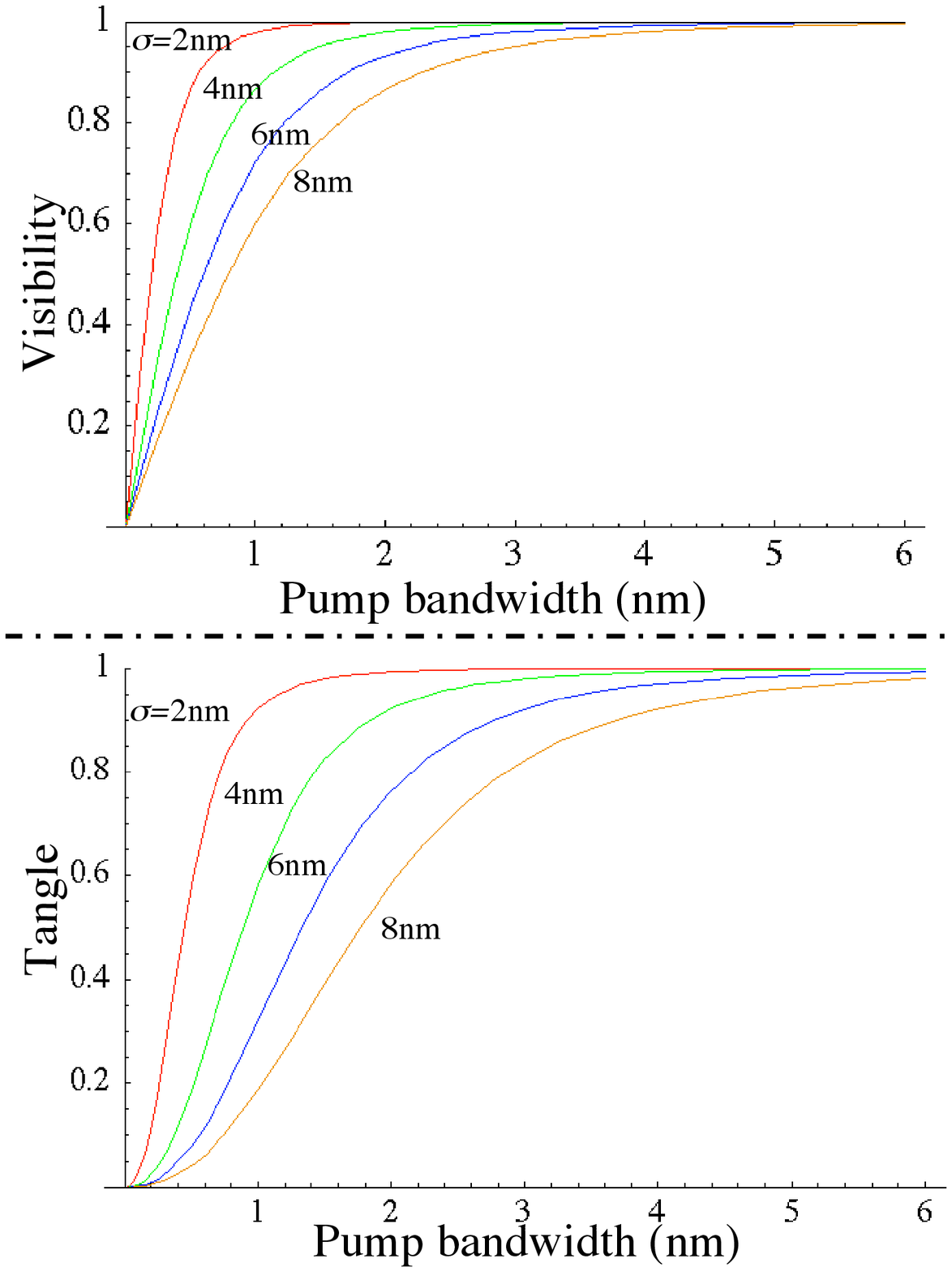}
\caption{(Color online) Comparison of the plots of the relative visibility and the tangle in the polarization state in eq. \eqref{massimo} as a function of the pump bandwidth in the case of a Gaussian pump. Different curves correspond to different values of the detection bandwidth.}
\end{figure}

\section{Departures from the ideal case.}

The results obtained above neglect the finite size of the crystal and the effect of phase dispersion of the PDC photons; therefore they remain valid in the limit of short crystal.

So far we have disregarded the fact that the crystals have finite thickness. If this is taken into account,
the state \eqref{wf} has to be modified as
\begin{equation}
\label{wfm}
\ket{\psi_{12}}{=}\int d\,\omega_{p}\,d\omega_1\,d\omega_2\, \, \delta(\omega_p-\omega_1-\omega_2) \Delta_\nu(\omega_p-\Omega_p,\omega_1-\omega_2) S_\nu(\omega_p-\Omega_p)a^\dag_1(\omega_1)a^\dag_2(\omega_2)\ket{0},
\end{equation}
where $\Delta_\nu(\nu_p,\nu_-)$ is the function describing non-perfect phase matching. The amplitude \eqref{AF} is modified by substituting $F_\nu(t)e^{-4\delta_0^2t'^2}$ with the convolution,
\begin{equation}
G_\nu(t,t')=\int ds\, ds' F_\nu(s)e^{-4\delta_0^2s'^2}\tilde\Delta_\nu(s-t,s'-t'),
\end{equation}
where $\tilde\Delta_\nu(t,t')$ is the two dimensional Fourier transform of $\Delta_\nu(\nu_p,\nu_-)$. A similar substitution should be carried out for the second pump beam.

In the case of degenerate type II phase matching, this can be approximated as \cite{rubin},
\begin{equation}
\label{sinc}
\Delta_\nu(\nu_p,\nu_-)=\text{sinc}(\alpha_p\nu_p+\alpha_-\nu_-),
\end{equation}
being $\alpha_p$ and $\alpha_-$ constants which depend on the length $L$ and the dispersion in the crystal, viz.
\begin{equation}
\begin{aligned}
&\alpha_p=\frac{L}{4}\left(\frac{1}{u_o(\Omega)}+\frac{1}{u_e(\Omega)}\right)-\frac{L}{2u_e(\Omega_p)},\\
&\alpha_-=\frac{L}{2} \left(\frac{1}{u_o(\Omega)}-\frac{1}{u_e(\Omega)}\right),
\end{aligned}
\end{equation}
where $u_o(\Omega)$ ( $u_e(\Omega)$) is the group velocity dispersion for an ordinary (extraordinary) beam at the frequency $\Omega$.
In this case, the expression for its transform is,
\begin{equation}
\tilde\Delta_\nu(t,t')=\frac{1}{\alpha_p}\text{rect}(\frac{t}{\alpha_p})\delta(t'-\frac{\alpha_-}{\alpha_p}t),
\end{equation}
which leads to the result,
\begin{equation}
G_\nu(t,t')=\frac{1}{\alpha_p}\int_{-\alpha_p/2}^ {\alpha_p/2}ds\,F_\nu(t+s)e^{-4\delta_0^2\left(t'+\frac{\alpha_-}{\alpha_p}s\right)^2} .
\end{equation}
Note that in the short crystal limit $\alpha_p\rightarrow0$, we restore the expression $G_\nu(t,t')=F_\nu(t)e^{-4\delta_0^2(t')^2}$.  In the opposite limit of infinite crystal the correlation \eqref{sinc} becomes perfect
\begin{equation}
\nu_p=-\frac{\alpha_-}{\alpha_p}\nu_-.
\end{equation}
Being $\alpha_p, \alpha_-$ linearly dependent on the crystal length $L$, the ratio $\frac{\alpha_-}{\alpha_p}$ remains finite when passing to this limit.  Such a correlation implies that a PDC pair with a frequency mismatch $\nu_-$ can be generated only by the component of the pump beam detuned of $-\frac{\alpha_p}{\alpha_-}\nu_p$.  Therefore, it can reduce the advantage of using a broad spectrum pump, since the crystal itself limit the matching bandwidth. As an example, in Fig. 5 the results  for degenerate type II PDC from a 400nm pump pulse are shown for two widely adopted materials: LiB$_3$O$_5$ (LBO), and BaB$_2$O$_4$ (BBO).  The pump bandwidth is 5nm, while the detection filters are chosen to have 2nm FWHM: in the ideal case, these values give near perfect visibility. It is clearly shown that the finite size of the crystal can be neglected for thicknesses in the range 2mm to 4mm. This is the size of bulk crystal slabs commonly adopted in experiments. At these wavelength, the adoption of LBO instead of BBO reduces the distinguishability effect. Anyway that would cause a significant reduction in the brightness of the source, given the smaller nonlinearity of LBO.

As a final remark, we note that when using  type II PDC it is convenient to use the signal photons as triggers and the idler photons in the interferometer. In fact, distinguishability of  signal and idler photons can arise from different dispersion in the PDC crystal due to birefringence.

\begin{figure}[!htb]
\includegraphics[width=0.49\columnwidth, bb=100 423 500 700,clip]{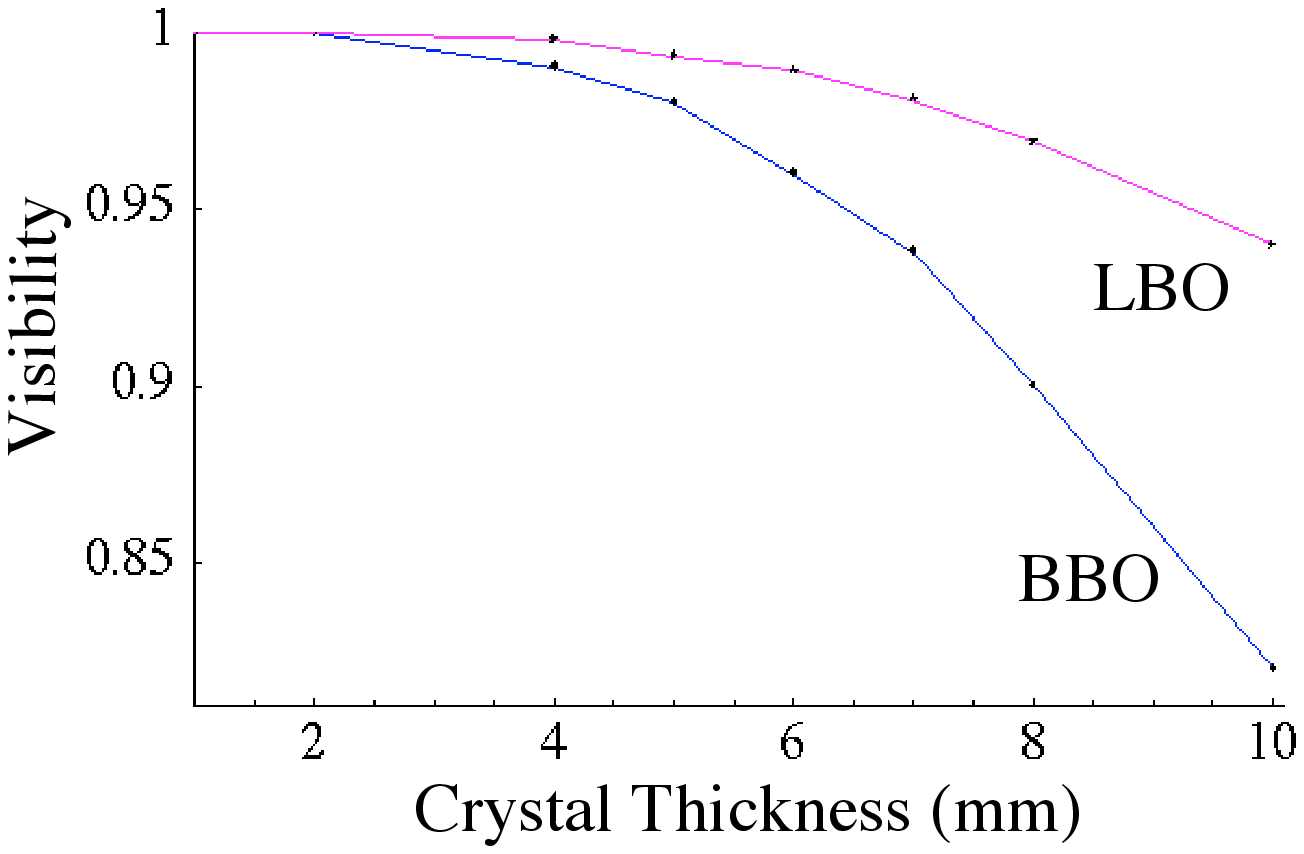}
\caption{(Color online) Plot of maximum achievable visibility as a function of crystal length in Type II degenerate 400nm$\rightarrow$800nm PDC, for two widely adopted materials, LBO and BBO. The pump bandwidth is 5nm, the detection bandwidth is 2nm. This choice gives near unity visibility in the ideal case. Sellmeier coefficients  are taken from Ref.\cite{handbook}}
\end{figure}

\section{Conclusion}
The results we have presented indicate that  when running entangling gates with independent PDC photons, large bandwidth pulses are somehow preferable to narrow ones. Their adoption reduces the {\em welcher weg} knowledge in the interference process coming from frequency entanglement. We showed that visibility has to be considered a forgetful parameter in the design of these experiments. The non trivial relation between tangle and visibility implies that the conditions to fulfill to attain a high level of entanglement are stricter than those for reaching near perfect visibilities. Such an effect of frequency correlation can be easily reduced by properly choosing pumping and detection conditions with existing materials.  Further investigation of this process can include the effects of group velocity dispersion on the arrival time of the PDC photons, resulting in a time jitter \cite{thomas}

At a first sight, it might seem counterintuitive that broadband pumping is better than single mode pumping; in fact it has been commonly observed that interference of dependent photons in pulsed regime is more difficult to reach than with cw pumping.  Nevertheless, using a mode locked pump {\em per se} does not present  in principle limitations. For the independent photon case, the actual temporal shape of the pump pulse is crucial, since knowledge comes from non trivial correlations between the pump pulses. In real life implementations, other sources of imperfections may impose restrictions on pump and detection bandwidth, e.g. dispersion in dielectric mirrors, wavelength-sensitive performance of beam splitters. Furthermore, restricting the duration of the pump pulses is less resilient to jitter and narrow detection filter reduce the apparent brightness of the source, leading to longer measurement times.

\section*{Aknowledgements}
We wish to acknowledge T.J. Weinhold, A. Gilchrist, A. G. White, T. A. Nieminen and T. Jennewein for useful discussions and comments. This work has been supported in part by the Australian Research Council,  and the US Disruptive Technologies Office.
\section*{Note added}
During the refereeing process of this manuscript, we become aware of a paper presenting  a similar treatment \cite{china}.


\begin{thebibliography}{99}

\bibitem{mandel} L. Mandel, {\em Phys. Rev. A} {\bf 28}, 929 (1983).

\bibitem{paul} H. Paul, {\em Rev. Mod. Phys.}, {\bf 58}, 209 (1986).

\bibitem{klm} E. Knill, R Laflamme, and G. Milburn, {\em Nature} (London) {\bf 409}, 46 (2001).

\bibitem{HOM} C. K. Hong, Z. Y. Ou, and L. Mandel, {\em Phys. Rev. Lett.} {\bf 59}, 2044-2046 (1987).


\bibitem{nc} M. A. Nielsen, and I. L. Chuang, {\em Quantum Computation and Quantum Information Theory }(Cambridge University Press,  Cambridge, UK, 2000).

\bibitem{briegel} R. Raussendorf and H. J. Briegel, {\em Phys. Rev. Lett.} {\bf 86}, 5188 (2001).

\bibitem{mike} M. A. Nielsen, {\em Phys. Rev. Lett.} {\bf 93}, 040503 (2004).

\bibitem{owexp} P. Walther {\em et al.}, {\em Nature} {\bf 434}, 169 (2005); R. Prevedel, {\em et al.}, ibidem {\bf 445}, 65 (2007).

\bibitem{clusterpan} C.-Y. Lu {\em et al.}, {\em Nature Phys.} {\bf 3}, 91 (2007).

\bibitem{clusterpino} G. Vallone,  E. Pomarico, P. Mataloni, F. De Martini, and V. Berardi, {\em Phys. Rev. Lett.} {\bf 98}, 180502 (2007).

\bibitem{source1} C. Santori, M. Pelton, G. Solomon, Y. Dale, and Y. Yamamoto, {\em Phys. Rev.
Lett.} 86, 1502 (2000).

\bibitem{source2} T. Wilk, S. C. Webster, H. P. Specht, G. Rempe, and A. Kuhn, {\em Phys. Rev. Lett.}  {\bf 98}, 063601 (2007).

\bibitem{qptcnot} J.L. O'Brien, {\em et al.}, {\em Phys. Rev. Lett.} {\bf 93}, 080502 (2004).  

\bibitem{cnot1} N. K.  Langford, et at., {\em Phys. Rev. Lett.} {\bf 95}, 210504 (2005).

\bibitem{cnot2} N. Kiesel, {\em et al.}, {\em Phys. Rev. Lett.} {\bf 95}, 210505 (2005).

\bibitem{cnot3} R. Okamoto, H. F. Hofmann, S. Takeuchi, and K. Sasaki, {\em Phys. Rev. Lett.} {\bf 95}, 210506 (2005).

\bibitem{panint} T. Yang, {\em et al.}, {\em Phys. Rev. Lett.} {\bf 96}, 110501 (2006) 

\bibitem{int}  R. Kaltenbaek, B. Blauensteiner, M. Zukowski, M. Aspelmeyer and A. Zeilinger, {\em Phys. Rev. Lett.} {\bf 96}, 240502 (2006).

\bibitem{till} T. J. Weinhold, {{\em et al.}}, in preparation.

\bibitem{goedel} B.-G. Englert, {\em Phys. Rev. Lett.} {\bf 77}, 2154 (1996). 

\bibitem{kwiat} P. D. D. Schwindt, P. G. Kwiat, and B.-G. Englert, {\em Phys. Rev. A} {\bf 60}, 4285-4290 (1999).
 
 \bibitem{fabio} F. Sciarrino, M. Ricci, F. De Martini, R. Filip, and L. Mista Jr, {\em Phys. Rev. Lett.}, {\bf 96}, 020408 (2006).

\bibitem{uren} W. P. Grice, A. B. U'Ren, and I. A. Walmsley, {\em Phys. Rev. A} {\bf 64}, 063815 (2001).

\bibitem{carrasco} S. Carrasco, {\em et. al., Opt. Lett.} {\bf 31}, 253-255 (2006).

\bibitem{eccala}   M. Zukowski, A. Zeilinger, and H. Weinfurter, {\em Ann. N.Y. Acad. Sci.} {\bf 755}, 91 (1995). 

\bibitem{eccala1} A. Zeilinger, M.A.  Horne, H. Weinfurter, and M. Zukowski, PRL, 78,  3031 (1997). 

\bibitem{rubin} T. E. Keller, and M. H. Rubin, {\em Phys. Rev. A}, {\bf 54}, 1534 (1997).



\bibitem{loudon} R. Loudon, {\em The Quantum Theory of Light} 3$^{rd}$ ed. (Oxford Press, Oxford, 2000).

\bibitem{glauber} R. Glauber, {\em Phys. Rev.} {\bf 130}, 2529 (1963); {\em ibidem} {\bf 131}, 2766 (1963).

\bibitem{note} In actual experiments, the coincidence window has to be also chosen shorter than the inverse repetition rate of the mode locked pump, in order to avoid accidental coincidence counts due to subsequent pulses.


\bibitem{wootters} W. K. Wootters, {\em Phys. Rev. Lett.} {\bf 80}, 2245 (1998).

\bibitem{handbook} V.G. Dmitriev, G.G. Gyrzadyan, and D.N. Nikogosyan {\it Handbook of Nonlinear Optical Crystals} (Springer-Verlag, Heidelberg, D, 1999).


\bibitem{thomas} T. Jennewein, private communication.

\bibitem{china} H.R. Zhang, and R. P. Wang, {\it Phys. Rev. A} {\bf 75}, 053804 (2007).


\end{thebibliography}
\end{document}